# Emergent excitation at the magnetic metal-insulator transition in the pyrochlore osmate $Cd_2Os_2O_7$

S. Calder,[1,*] J. G. Vale,[2] N. A. Bogdanov,[3] X. Liu,[4,5] C. Donnerer,[2] M. H. Upton,[6] D. Casa,[6] M. D. Lumsden,[1] Z. Zhao,[7,8] J.-Q. Yan,[7,8] D. Mandrus,[7,8] S. Nishimoto,[3] J. van den Brink,[3] J. P. Hill,[4] D. F. McMorrow,[2] and A. D. Christianson[1,9]

[1]Quantum Condensed Matter Division, Oak Ridge National Laboratory, Oak Ridge, TN 37831, USA.
[2]London Centre for Nanotechnology, University College London, London, WC1H 0AH, UK.
[3]Institute for Theoretical Solid State Physics, IFW Dresden, D01171 Dresden, Germany.
[4]Condensed matter physics and materials science department, Brookhaven National Laboratory, Upton NY 11973, USA.
[5]Institute of Physics, Chinese Academy of Sciences, Beijing, China.
[6]Advanced Photon Source, Argonne National Laboratory, Argonne, IL 60439, USA.
[7]Department of Materials Science and Engineering, University of Tennessee, Knoxville, TN 37996, USA.
[8]Materials Science and Technology Division, Oak Ridge National Laboratory, Oak Ridge, TN 37831, USA.
[9]Department of Physics & Astronomy, University of Tennessee, Knoxville, TN 37996, USA.
*Correspondence to: caldersa@ornl.gov

**The rich physics manifested by 5d oxides falls outside the Mott-Hubbard paradigm used to successfully explain the electronic and magnetic properties of 3d oxides. Much consideration has been given to the extent to which strong spin-orbit coupling (SOC), in the limit of increased bandwidth and reduced electron correlation, drives the formation of novel electronic states, as manifested through the existence of metal-insulator transitions (MITs). SOC is believed to play a dominant role in $5d^5$ systems such as iridates ($Ir^{4+}$), undergoing MITs which may or may not be intimately connected to magnetic order, with pyrochlore and perovksite systems being examples of the former and latter, respectively. However, the role of SOC for other 5d configurations is less clear. For example, $5d^3$ (e.g $Os^{5+}$) systems are expected to have an orbital singlet and consequently a reduced effect of SOC in the groundstate. The pyrochlore osmate $Cd_2Os_2O_7$ nonetheless exhibits a MIT intimately entwined with magnetic order with phenomena similar to pyrochlore iridates. Here we report the first resonant inelastic X-ray scattering (RIXS) measurements on an osmium compound, allowing us to determine the salient electronic and magnetic energy scales controlling the MIT in $Cd_2Os_2O_7$, which we benchmark against detailed quantum chemistry calculations. In particular, we reveal the emergence at the MIT of a magnetic excitation corresponding to a superposition of multiple spin-flip processes from an Ising-like all-in/all-out magnetic groundstate. We discuss our results with respect to the role of SOC in magnetically mediated MITs in 5d systems.**

The diverse physics of transition metal oxides has stimulated interest for decades. Particular focus has resided on 3d oxides where the strong electron correlations dominate, with a dramatic manifestation being the occurrence of the Mott MIT. [1, 2] Contrastingly in 5d oxides the relativistic SOC is increased to such an extent that competition occurs with the on-site electron correlations, as well as further interactions such as the crystalline electric field (CEF), local multiplet physics and bandwidth. The consequence of these often finely balanced interactions in 5d oxides is the emergence of new physics, such as a SOC dominated Mott-like insulating state, initially observed in perovskite iridates, [3, 4] and Weyl semi-metal, non-trivial topological insulators and magnetic MITs in pyrochlore iridates. [2, 5] We focus here on the pyrochlore osmate $Cd_2Os_2O_7$ where the concomitant magnetic ordering and MIT cannot be reconciled with the Mott-Hubbard paradigm. [6, 7, 8, 9, 10] Instead the behavior was initially considered to be the first manifestation of a Slater transition where the onset of magnetic order creates the insulating phase by introducing a periodic potential that localizes the conduction electrons. [11] More recently, however, the mechanism of the MIT has been argued to be a Lifshitz transition normally associated with metal-metal transitions and generally not explicitly requiring magnetic order. [6, 7]

Phenomenologically the behavior of $Cd_2Os_2O_7$ is analogous to iridate pyrochlores: they both undergo temeperature dependent continuous MITs suggested to be associated with all-in/all-out (AIAO) magnetic ground states where all spins either point in or out of the center of the tetrahedron. [12, 13, 14] One consequence of the AIAO ground state is that since the ordering breaks time-reversal symmetry but maintains the cubic symmetry of the lattice such systems can host Weyl semi-metal behavior. [5] In this investigation we are able to access the magnetic Os sublattice with neutrons and confirm the AIAO magnetic structure, thereby going beyond the strong but indirect evidence presented for AIAO ordering of the Ir sublatttice in the pyrochlore iridates. Although there exists apparent similarities between osmate and iridate pyrochlores they contain distinct electronic occupancies of $5d^5$ in the iridates and $5d^3$ in osmates. Consequently SOC has been considered to play an explicit role in the behavior of iridates while in the osmates the influence of SOC is less clear and often considered implicitly. By probing the electronic and magnetic ground and excited states in $Cd_2Os_2O_7$ we are able to disintangle the degree to which SOC determines the manifested behavior and in doing so consider both the similarities and distinctions between $5d^3$ and $5d^5$ based systems.

In order to access the exotic physics in $Cd_2Os_2O_7$, and 5d systems in general, it is crucial to explicate the electronic ground state and subsequent excited states that emerge. These derive from the competing interactions acting on the 5d ion and consequently the relevant microscopic interactions, and their energy hierarchy, need to be determined. By performing momentum and temperature dependent osmium L-edge RIXS we directly probe the 5d electrons and uncover the salient features of the electronic ground state and magnetic excitations. These measurements extend the RIXS technique to 5d materials beyond iridates, [15, 16, 17, 18, 19, 20] and in doing so provide access to previously inaccesible spin excitations. To guide our Os RIXS measurements and gain fundamental insights we make contact with detailed theory. In $Cd_2Os_2O_7$ we find electronic excitations that are markedly different from iridates thereby revealing a divergent role of SOC between these systems, particularly in the degree that SOC plays in the creation of the electronic ground state. However, when considering interactions that derive from more than one Os ion we find an elevated role of SOC in $Cd_2Os_2O_7$ and similar behavior to pyrochlore iridates as, for example, manifested by the magnetic ground state that points towards related mechanisms driving the MITs in both systems. Moreover, our RIXS measurements reveal an excitation characteristic of magnon energies. By considering the relevant energy scales and interactions uncovered allows a comprehensive description of this excitation as a magnetic excitation that emerges in $Cd_2Os_2O_7$ due to the cooperation of the large SOC, single ion-anisotropy and Dzyaloshinskii-Moria (DM) interactions that combine to form a superposition of spin states.

**Results**

**RIXS measurements at the osmium $L_3$-edge within the magnetic insulating phase of $Cd_2Os_2O_7$.** We first discuss fixed incident energy RIXS measurements, the spectrum is shown in Fig. 1**a**. Two pronounced features are evident, labeled "$E_B$" and "$E_C$", each significantly broader than the experimental energy resolution and located at $E_B$=0.92(6) eV and $E_C$=4.5(1) eV. In addition, a small, sharp resolution-limited feature "$E_A$" is observed at $E_A$=0.16(1) eV. ($E_A$ is more apparent in Fig. 2). Much can be learned by probing the intensity dependence of the inelastic spectra at different fixed incident energies, as shown in Fig. 1**b**. These measurements reveal features $E_A$ and $E_B$ have their maximum resonant intensity at the same incident energy, 10.8755(5) keV, whereas $E_C$ shows a maximum intensity at a higher incident energy of 10.879(1) keV. The different incident energy dependence is a consequence of the excitations accessing different core-hole (2p-5d) transitions during L-edge RIXS due to the splitting of the osmium 5d manifold, nominally into $t_{2g}$ and $e_g$ sub-manifolds. [17, 18] The scattering involving excitations within the $t_{2g}$ manifold will occur at a lower energy than scattering involving $e_g$ levels, with the energy difference corresponding to the $t_{2g}$ and $e_g$ splitting. This allows us to assign features $E_A$ and $E_B$ to intra-$t_{2g}$ processes and $E_C$ as involving an $e_g$ process.

To further interpret the underlying processes leading to the measured RIXS spectra we benchmark our results against recent many-body quantum chemistry calculations. [21] Those calculations predicted Os d-d multiplet excitations in $Cd_2Os_2O_7$ starting around 1.5 eV, which is consistent with the measured energy for $E_B$ and moreover corresponded to an intra-$t_{2g}$ process. This intra-$t_{2g}$ ($t_{2g}^3 \rightarrow t_{2g}^3$) excitation at $E_B$ corresponds to a spin flip of one of the three electrons in the $5d^3$ valance band at the osmium site changing the total local spin from high-spin S=3/2 to low-spin S=1/2, shown in Fig. 1**c**. The energies of this excitation are controlled by the Hund's coupling ($J_H$) that in the atomic limit are $3J_H$ and $5J_H$ (see e.g. Ref. 22), giving an estimate of $J_H \approx 0.3$ eV in $Cd_2Os_2O_7$. This type of Hund ("spin") excitation, which has been observed in 3d oxides, has never been previously measured in a 5d system. The presence of $E_B$ immediately reveals a different d-manifold electronic ground state in $Cd_2Os_2O_7$ compared to the Mott-like iridates. Specifically, Ir L-edge RIXS measurements show appreciable SOC splitting of the $t_{2g}$ manifold in the form of a "spin-orbit" exciton, [15, 16, 17, 18, 19] for example in the pyrochlore iridates the measured $t_{2g}$ splitting is of order 0.4 eV. [23] The absence of a SOC driven exciton in the RIXS spectra of $Cd_2Os_2O_7$, with instead the presence of $E_B$, indicates that SOC does not strongly split the $t_{2g}$ manifold, i.e. the SOC driven $J_{eff}=1/2$ electronic ground state is not realized in this $5d^3$ osmate.

The higher energy excitation $E_C$ involves transitions that access the $e_g$ level, as shown in Fig. 1**b**. Calculations predicted inter $t_{2g}$-$e_g$ excitations ($t_{2g}^3 \rightarrow t_{2g}^2 e_g^1$) at 5 eV,[16] again about 0.5 eV above the observed RIXS peak. The divergence between predicted and observed excitations, at least in part, can be explained by the details of the calculations in Ref. 21, where altering the number of states or description of the nearest neighbor sites in the spin-orbit treatment will affect the predicted energies. Excitation $E_C$, shown in Fig. 1**c**, is a direct measurement of the 10Dq CEF splitting of 4.5 eV in $Cd_2Os_2O_7$. For SOC to play a role in the ground state electronic configuration it needs to intermix the $t_{2g}^3$ manifold itself, which our measurement of feature $E_B$ indicates does not appreciably occur, or mix it with $t_{2g}^2 e_g^1$ states. However, with a large 10Dq splitting of 4.5 eV, the latter mixing is strongly prohibited. Conversely in the pyrochlore iridates where SOC does impact the ground state the CEF splitting is around 1 eV lower than in $Cd_2Os_2O_7$.[23]

Collectively the RIXS results for d-d transitions in $Cd_2Os_2O_7$ reveal an electronic ground state of the $5d^3$ ion in $Cd_2Os_2O_7$ that is dominated by CEF (4.5 eV) followed by Hund's coupling (0.3 eV). The much weaker role of SOC in the creation of the electronic ground state of the $Os^{5+}$ ion, counter intuitively, is not at odds with the expectations of observable effects of the large SOC that is intrinsic to 5d systems.[6, 10] For example, as we discuss further, the role of SOC comes to the fore when either considering excited magnetic and electronic states or when going beyond the single ion ground state.

**Momentum and temperature dependence of excitation $E_A$ with RIXS.** Having characterized the electronic ground state from RIXS excitations at 1 eV and above we now focus on feature $E_A$ at 160 meV. This energy does not correspond to any expected "d-d" energy scale for the $d^3$ electronic configuration in a nearly cubic CEF, see e.g. Ref. 24. Moreover, $E_A$ is distinct from the d-d excitations $E_B$ and $E_C$ in being much sharper in energy. Figure 2**a-c** shows the intensity of $E_A$ follows an order parameter-like behavior with temperature, with $E_A$ appearing at the magnetic MIT, and remains at the same energy of 160 meV. We followed the momentum dependence, main panel of Fig. **2d**, and conclude $E_A$ is a non-dispersive excitation, within the 130 meV experimental resolution. While the intensity of $E_A$ appears to show some variation within the Brillouin zone, this is, at least in part, an artifact of the observed variation of the elastic line as the crystal is necessarily measured in slightly different physical orientations altering the x-ray beam attenuation.

The origin and behavior of mode $E_A$ is puzzling for a variety of reasons. Interpretation of this feature in terms of a conventional magnetic excitation initially appears problematic given the small calculated value of the nearest-neighbour exchange interaction J=6.4 meV for $Cd_2Os_2O_7$.[16] Therefore we first consider potential non-magnetic mechanisms. One scenario, given the concomitant MIT, is a relationship with the insulating state. Such an excitation was observed in the iridate $A_2IrO_3$,[19] with an excitation of 340 meV corresponding exactly to the Mott gap size. A similar origin, however, is inconsistent with known behavior of $Cd_2Os_2O_7$, for which the insulating gap has been shown to open continuously, a fact that has been cited in favour of a Slater mechanism.[7,8] Therefore, if feature $E_A$ were related to the insulating gap, one would expect to observe a significant shift in the energy with temperature. This is not the case. Moreover, the insulating gap, $2\Delta$, in $Cd_2Os_2O_7$ is around 100 meV[8] and is therefore not consistent with the measured RIXS spectra. We finally rule out a d-d excitation scenario by noting that the $t_{2g}$ levels are already split at high temperature (>225 K) in $Cd_2Os_2O_7$, indeed this is one of the causes of the single-ion anisotropy in the system. Any alteration of the $t_{2g}$ splitting due to lower symmetry crystal fields below the magnetic MIT will not produce such a low-lying d-d excitation. This is supported by the RIXS measurements not observing any change in $E_B$ or $E_C$ that would indicate an altered ground state and potential routes for $E_A$.

**Magnetic origin of $E_A$ from an all-in/all-out ground state.** We argue that $E_A$ does indeed have a magnetic origin and support this with numerical exact diagonalisation (ED) calculations. The magnetic ground state in $Cd_2Os_2O_7$ has been shown to be consistent with the AIAO ordering,[7] however as with the predictions of AIAO ordering on the pyrochlore iridates this relied on some conjecture based on structural arguments. To test this magnetic ordering picture in $Cd_2Os_2O_7$, we carried out neutron diffraction measurements. The measurements revealed magnetic Bragg peaks that confirmed the proposed AIAO magnetic structure and furthermore allowed the ordered moment to be found of 0.59(8)$\mu_B$/Os, see Fig. 3**a-b**. The AIAO magnetic structure is stabilized in the frustrated pyrochlore lattice with strong single-ion axial anisotropy.[6, 7, 21] Therefore in the leading approximation the spins can be considered to be Ising-like local S=3/2. In this case the resultant classical Hamiltonian that includes easy-axis anisotropy (D), Heisenberg exchange (J) and the DM interaction (d) gives three types of local spin excitations for this S=3/2 system of $S_z$=3/2→1/2 ($\Delta S_z$=1), $S_z$=3/2→-1/2 ($\Delta S_z$=2) and $S_z$=3/2→-3/2 ($\Delta S_z$=3). These have the following energies:

$$\Delta S_z=1: \Delta E=(\sqrt{2}*d + J/2)N - D$$

$$\Delta S_z=2: \Delta E=(2\sqrt{2}*d + J)N - D \qquad (1)$$

$$\Delta S_z=3: \Delta E=(3\sqrt{2}*d + 3J/2)N$$

To explore the relevance of the $\Delta S_z$ processes to our RIXS data we performed ED calculations. The ground state and all possible excited states of the Hamiltonian

$$H = J\sum_{<i,j>} \vec{S}_i \cdot \vec{S}_j + D \sum_i (\vec{S}_i \cdot \vec{A}_i)^2 + d \sum_{<i,j>} \vec{e}_{ij}[\vec{S}_i \times \vec{S}_j] \qquad (2)$$

with unitary vectors $\vec{A}_i \in \langle 111 \rangle$ and $\vec{e}_{ij} \in \langle 110 \rangle$ were obtained for given 4-site and 8-site clusters that encompass the full AIAO magnetic ground state with a fixed parameter set reported in Ref. 16, that includes d=1.7 meV, D=-6.8 meV and J=6.4 meV. The ED calculations take into account explicitly the quantum nature of the spin states and the interactions between them. The results are shown in Fig. 3**c**. The calculated energy shows excellent agreement with experiment. We note that the intensities calculated are the density of states (DOS) and therefore are not directly proportional to the RIXS cross-section, which is non-trivial to calculate.

In the classical limit, equation 1, the possible spin-flip excitations for a S=3/2 system of $\Delta S_z$=1,2,3 have distinct energies. Conversely the spectra for the $\Delta S_z$ excitations from ED calculations are mixed and rather similar with an overlapping single peaked excitation. The only distinction between the different $\Delta S_z$ processes from ED calculations is a slight variation in energy and an overall intensity-scaling factor in their DOS, as shown by the red, green and blue regions in Fig. 3**c**. The calculated mixing is due to quantum fluctuations leading to a superposition of different spin-states that occurs due to the DM interaction being appreciable in strength in $Cd_2Os_2O_7$, J/d=3, which is in turn related to the strong SOC intrinsic in $Cd_2Os_2O_7$. The DM interaction mixes states with different spin projection $S_z$ into both the ground state and the excited states. This mixing occurs to such an extent that the separate $\Delta S_z$=1,2,3 excitation channels become indistinguishable and result in spectra with broad tails and a maximum of intensity very close to the energy measured for $E_A$ of 160 meV.

The strong agreement between the observed magnetic excitation energy and that predicted from ED calculations based on the inclusion of the J, d and D interactions predicted in Ref. 16, with essentially no free parameters, indicates that this model robustly describes the essential physics of the system. Interestingly the ED calculations predict the strongest DOS contribution to be from the $\Delta S_z$=3 process. Such an excitation is forbidden by RIXS spin-only selection rules that limits the possible measurable excitations to $\Delta S_z$=1 and 2. [25] The $\Delta S_z$=3 process, however, becomes experimentally allowed in $Cd_2Os_2O_7$ in the intermediate RIXS process due to the creation of a $5d^4$ state ($2p^65d^3 \rightarrow 2p^55d^4$). In this intermediate state S is no longer a good quantum number, nevertheless, the experimental RIXS cross section will be expected to be dominated by the single-magnon followed by the two-magnon processes.

Further experimental support for the magnetic origin of excitation $E_A$ is found when recalling the RIXS incident energy dependence (Fig. 1**b**). The incident energy spectra showed both $E_A$ and $E_B$ had the same incident energy resonance dependence indicating they both involve solely intra-$t_{2g}$ excitations. Additionally, we observed $E_A$ is non-dispersive (Fig. 2**d**). This is indeed expected in a system that is predominantly of Ising type since the AIAO structure is a lowest energy, rather than degenerate, ground state, and will suppress the propagation of a flipped spin that alters the magnetic ordering.

**Discussion**

The Os L-edge RIXS and neutron measurements have provided direct access to the 5d electrons, competing inter and intra-ion interactions and AIAO magnetic ground state and subsequent excitations in $Cd_2Os_2O_7$. This allows a detailed and general understanding of the role of SOC and magnetism in creating the MITs in both osmate ($5d^3$) and iridate ($5d^5$) pyrochlores where the behavior is phenomenologically similar but the electronic ground state is revealed as distinctly different. In the $5d^5$ iridates there is one hole in the $t_{2g}$ shell giving rise to three (nearly) degenerate orbital configurations carrying orbital moment $l_{eff}=1$ and spin $S=1/2$, which SOC splits into a $J=1/2$ doublet and a $J=3/2$ quartet. This splitting is measureable with RIXS as a well-defined spin-orbit exciton [15, 16, 23] Conversely in $Cd_2Os_2O_7$ there are three valence electrons that half-fill the $t_{2g}$ shell, with spins aligned due to Hund's rule that consequently creates a further interaction in $5d^3$ systems that is negated in the iridates with $5d^5$. Therefore, to the leading order $Cd_2Os_2O_7$ carries a spin $S=3/2$ and no orbital moment $l_{eff}=0$. The orbital moment is quenched because there is only one possible orbital configuration with three spins aligned. The RIXS spectra of $Cd_2Os_2O_7$ provides direct evidence of this via the measurement of excitation $E_B$ that reveals a $t_{2g}$ manifold with no observable splitting. This occurs since SOC cannot affect the half-filled $t_{2g}$ shell, apart from second order effects. In this strict sense SOC plays a negligible role in creating the electronic $S=3/2$ ground state in $5d^3$ systems. Considering second order effects in $Cd_2Os_2O_7$ the leading correction comes from the fact that the $e_g$ states, while well separated, are not infinitely far above the $t_{2g}$ shell. SOC can therefore admix $e_g$ states into the $t_{2g}$ $S=3/2$ ground state. This effect is governed by the ratio of the SOC $\lambda$ (~0.4 eV) [5] and the crystal field splitting 10Dq (~4.5 eV). The leading order result of this mixing is a single ion magnetic anisotropy: a splitting between the $S_z=+/-1/2$ and the $S_z=+/-3/2$ projections of the $S=3/2$ manifold, where z is the easy axis of the system. For $Cd_2Os_2O_7$ this results in an easy axis anisotropy of around 7 meV that determines the magnetic ground state.[16] Further consequences of the SOC exist when going beyond single-site considerations, such as the DM interaction. Therefore, while SOC does not dominate the mechanisms creating the electronic ground state of a single Os ion, in contrast to the Mott-like iridates, it still strongly impacts the physics of $Cd_2Os_2O_7$.

The large SOC in $Cd_2Os_2O_7$, which is to a large extent quenched in the electronic ground state, manifests in the stabilization of the AIAO magnetic ground state, as verified with neutron diffraction measurements. This non-degenerate ground state is selected in the frustrated pyrochlore structure due to the appreciable single-ion anisotropy in the system. A similar mechanism is expected to exist in pyrochlore iridates, although the AIAO ordering has not been directly measured on the Ir sublattice. Nevertheless the AIAO magnetic ordering has been the focus of considerable interest in pyrochlore iridates due to the potential for exotic phenomena including Weyl semi-metal and topological insulating behavior.[2, 5, 12, 13, 14] Analogous behavior can be mapped over to pyrochlore osmates with AIAO ordering. In terms of the observed MITs in pyrochlore iridates the AIAO ordering is predicted to play a direct role, with the existence of either concurrent or proximate magnetic ordering.[5] Considering the phenomenologically similar behavior between the pyrochlore iridates and the pyrochlore osmate $Cd_2Os_2O_7$ suggests an analogous underlying mechanism for the magnetic MITs. Our results indicate neither of the divergent electronic ground states adopted within the $5d^3$ osmate and $5d^5$ iridate pyrochlore systems of orbital singlet with quenched SOC and SOC enhanced $J_{eff}=1/2$, respectively, play a dominant role with regards to creating the MITs. Instead the underlying mechanism appears directly related to the enhanced SOC in both systems that produces the strong single-ion anisotropy and subsequent AIAO magnetic ordering.

Considering the emergence of the magnetic excitation at $E_A=160$ meV in the RIXS measurements of $Cd_2Os_2O_7$ provides further evidence for the enhanced role of SOC in $5d^3$ systems. The existence of this excitation from the AIAO ground state was shown to be a direct consequence of the strong DM exchange interaction and single-ion axial anisotropy that couple with the comparable energy of the magnetic exchange interaction; with the concomitant behavior necessary to create the observed magnetic excitation. The coupling of these interactions on the Ising-like AIAO ground state results in a magnetic excitation that corresponds to a superposition of multiple spin-flip process that obeys the Hamiltonian described in equation 2, which includes comparable DM, single-ion axial anisotropy and magnetic exchange interactions.

Collectively the Os L-edge RIXS measurements on $Cd_2Os_2O_7$, coupled with neutron diffraction, have provided a unique and direct probe of the 5d electrons responsible for the concurrent magnetic order and MIT in both the metallic and insulating regimes. The results revealed the electronic ground state, showing a mechanism of strong CEF (10Dq≈4.5 eV), moderate Hunds coupling ($J_H$≈0.3eV) and quenched SOC that does not split the $t_{2g}$ manifold but enhances the anisotropic magnetic couplings to create an AIAO magnetic ground state. This reveals a non-trivial and variable role of effective SOC in $5d^3$ based systems compared to $5d^5$ systems, such as the $Ir^{4+}$ iridates, where SOC plays a more consistently dominant role. The behavior in $Cd_2Os_2O_7$ suggests that in general a route beyond the Mott-Hubbard paradigm is required to open the insulating gap in 5d pyrochlores with AIAO magnetism the likely driving mechanism.

## Methods

**Synthesis.** Single crystals of $Cd_2Os_2O_7$ of approximate dimensions $0.2mm^3$ were grown as described in Ref. 8 by sealing appropriate quantities of CdO, Os, and $KClO_3$ in a silica tube and heating at 800°C for one week. Prior to the RIXS measurement several crystals were characterized and aligned with an x-ray Laue. All RIXS measurements were performed on a single crystal of $Cd_2Os_2O_7$. Polycrystalline $Cd_2Os_2O_7$ was prepared with isotopic $^{114}Cd$ for neutron measurements to negate the extremely high neutron absorption of standard Cd using sold state techniques from $^{114}CdO$ and $OsO_2$ powders.

**Resonant inelastic X-ray scattering.** Single crystals of $Cd_2Os_2O_7$ were measured with RIXS at the Os $L_3$-edge (10.88 keV) on Sector 30 at the Advanced Photon Source (APS) using the MERIX instrumentation.[26] The incident energy was accessed with two pre-sample monochromators, a primary Si(111) monochromator and a secondary Si(444) monochromator. The energy of the beam scattered from the sample was discriminated with a Si(466) 2m diced analyzer. The detector was a MYTHEN strip detector. To reduce the elastic line we performed inelastic measurements in horizontal geometry within a few degrees of 90°. The RIXS energy resolution was 130meV FWHM. This compares favorably to initial work in the iridates and further efforts are aimed at achieving resolution below 50 meV at the Os $L_3$-edge. The $Cd_2Os_2O_7$ single crystal was oriented in the (HKK) scattering plane to access high symmetry directions in the Brillouin zone. Measurements were performed in the same Brillouin zone and all measurements were performed more than once. The fitting described in the text was based on a least squares analysis. For the dispersion relation in Fig. 2 a nominal error of 50 meV has been reported, that is larger than the error obtained by least squares, to account for the variation of the elastic line.

**Neutron powder diffraction measurements.** Neutron powder diffraction measurements were performed on the HB-2A powder diffractometer at the High Flux Isotope Reactor (HFIR), Oak Ridge National Laboratory. Isotopic $^{114}Cd_2Os_2O_7$ was used to overcome the substantial neutron absorption from elemental cadmium. Measurements were performed using a wavelength of 2.41 Å between 4 K and 250 K. The magnetic structure was refined using Fullprof and SARAh utilizing an irreducible representational analysis approach. The $\Gamma_3$ irreducible representation was used for the final solution, after trying all possible irreducible representation solutions for a propagation vector of k=(0,0,0) and finding this to be the only solution that described the data. The solution corresponds to the "all-in/all-out" magnetic structure for the magnetic Os ion at the (0,0,0) site in the pyrochlore structure with a propagation vector of k=(0,0,0). We obtained the magnetic moment from a Fullprof refinement by normalizing the magnetic scattering to the nuclear reflections and further checked the lower and upper error bounds were an accurate reflection of the data.

**Full exact diagonalisation calculations.** The ground state and all possible excited states of the Hamiltonian (equation 2) with unitary vectors $\vec{A}_i \in \langle 111 \rangle$ and $\vec{e}_{ij} \in \langle 110 \rangle$ were obtained for given 8-site and 4-site clusters with a parameter set reported in Ref. 16. Using the states we calculated the excitation spectrum $I(\omega)=\Sigma_n |\langle y_n|\hat{O}|y_0\rangle|^2 d(\omega-E_n+E_0)$, where $E_n$ are the n-th eigenstate and eigenenergy (n=0 corresponds to the ground state). The applied operator was taken as $\hat{O}=S_+, (S_+)^2, (S_+)^3$ for $\Delta S_z=1,2,3$ excitation channels, respectively. Additional calculations were run on a 4-site cluster, which is the minimum to reproduce the all-in/all-out magnetic structure, to test for finite size effects. The results between the 4-site and 8-site showed very little difference.

**Acknowledgments:** Work at ORNL's High Flux Isotope reactor was supported by the scientific User Facilities Division, Office of Basic Energy Sciences, U.S. Department of Energy (DOE). Use of the Advanced Photon Source, an Office of Science User Facility operated for the U.S. DOE Office of Science by Argonne National Laboratory, was supported by the U.S. DOE under Contract No. DE-AC02-06CH11357. Work in London was supported by the EPSRC. Work performed at Brookhaven National Laboratory was supported by the US DOE under contract no. DE-AC02-98CH10886.

**Author contributions:** S.C. and A.D.C conceived the investigation. J.G.V, X.L, C.D, M.H.U, D.C, A.D.C performed the RIXS experiment, as well as conceiving the experiment along with J.P.H, D.F.M, J.vd.B and M.D.L. S.C performed the neutron experiment. Z.Z, J.Y and D.M prepared the samples. N.B, S.N and J.vd.B. performed the theoretical calculations. S.C. led the manuscript preparation and all authors contributed.

**Figure legends**

**Figure 1 | RIXS measurements of $Cd_2Os_2O_7$ at the osmium L-edge. (a)** Fixed incident energy photons of $E_i$=10.877 keV, corresponding to the osmium resonant $L_3$-edge, probe an inelastic energy loss spectrum out to 7.5 eV (black dots). Three inelastic features are resolvable, $E_A$=0.16(1) eV, $E_B$=0.92(6) eV and $E_C$=4.5(1) eV. The horizontal bars indicate the instrument resolution of 130 meV (FWHM). The full RIXS spectrum was modeled with three Gaussians fitting the inelastic peaks $E_A$, $E_B$ and $E_C$, and a Gaussian for the elastic signal on top of a sloping background (blue line). **(b)** Varying the incident energy to follow the intensity dependence of the inelastic spectra reveals the maximum intensity of $E_A$ and $E_B$ at the same incident energy of $E_i$=10.8755(5) keV (solid line), whereas $E_C$ has a maximum at $E_i$=10.879(1) keV (dashed line). The difference of the resonant energies reflects the splitting of the osmium d-manifold and allows a categorization of the excitations as intra-$t_{2g}$ or $t_{2g}$-$e_g$. **(c)** Schematic of the initial and final RIXS process. The initial electronic ground state prior to exciting an electron is indicated by the red spins in the limit of cubic CEF splitting of the 5d manifold. The final RIXS states for $E_B$ (intra-$t_{2g}$) and $E_C$ ($t_{2g}$-$e_g$) are indicated by the blue and green spins, respectively. All measurements were performed at 60 K.

**Figure 2 | Temperature and momentum dependent RIXS. (a)** RIXS spectra at constant q=(2.5,8.8,8.8) and fixed incident energy $E_i$=10.877 keV at several temperatures through the magnetic MIT in $Cd_2Os_2O_7$. **(b)** The low energy excitation ($E_A$=160 meV), along with the elastic line, was fit to a Gaussian on a background from the higher energy scattering. **(c)** The intensity of $E_A$ increases below the magnetic MIT, the line is a fit to a power law. **(d)** RIXS measurements in the magnetic insulating regime performed along high symmetry directions in the magnetic Brillouin zone. The elastic scattering was suppressed by measuring within 4° of 2θ=90°, the remaining elastic signal has been subtracted from the data. (Inset top) The inelastic energy and intensity dependence of $E_A$ reveal dispersionless behavior for $E_A$. (Inset bottom) The Brillouin zone in the (HKK) plane is shown (grey) and the directions measured (red).

**Figure 3 | Magnetic excitation from an AIAO ground state in $Cd_2Os_2O_7$. (a)** Neutron powder diffraction measurements on isotopic $^{114}Cd_2Os_2O_7$, used to overcome the significant neutron absorption of Cd. The data (red circle) show the difference between measurements taken at 240 K and 160 K and was fit with an "all-in/all-out" magnetic structural model (black line). (Inset) The (220) magnetic reflection intensity, $2\Theta=39.1°$, shows $T_N=225$ K. **(b)** The AIAO magnetic structure formed in $Cd_2Os_2O_7$. **(c)** 8 site ED calculations, including resolution broadening, showing the character and amount of excited eigenstates at a particular energy. The scaled experimental data (black) is shown after background subtraction. The ED results reveal $\Delta S_z=1,2,3$ excitations (blue, green, red, respectively) with the strong DM interactions in the magnetic ground state and excited states of $Cd_2Os_2O_7$ resulting in a superposition of the spin states contributing to the excitation. The calculated intensity corresponds to the density of states and therefore the actual experimental intensity distribution of the excitation measured by the RIXS cross-section is likely dominated by the $\Delta S_z=1$ process. Inset corresponds to 4-site calculations and highlights the mixing and energy distribution of the $\Delta S_z$ excitations.

**Figure 1**

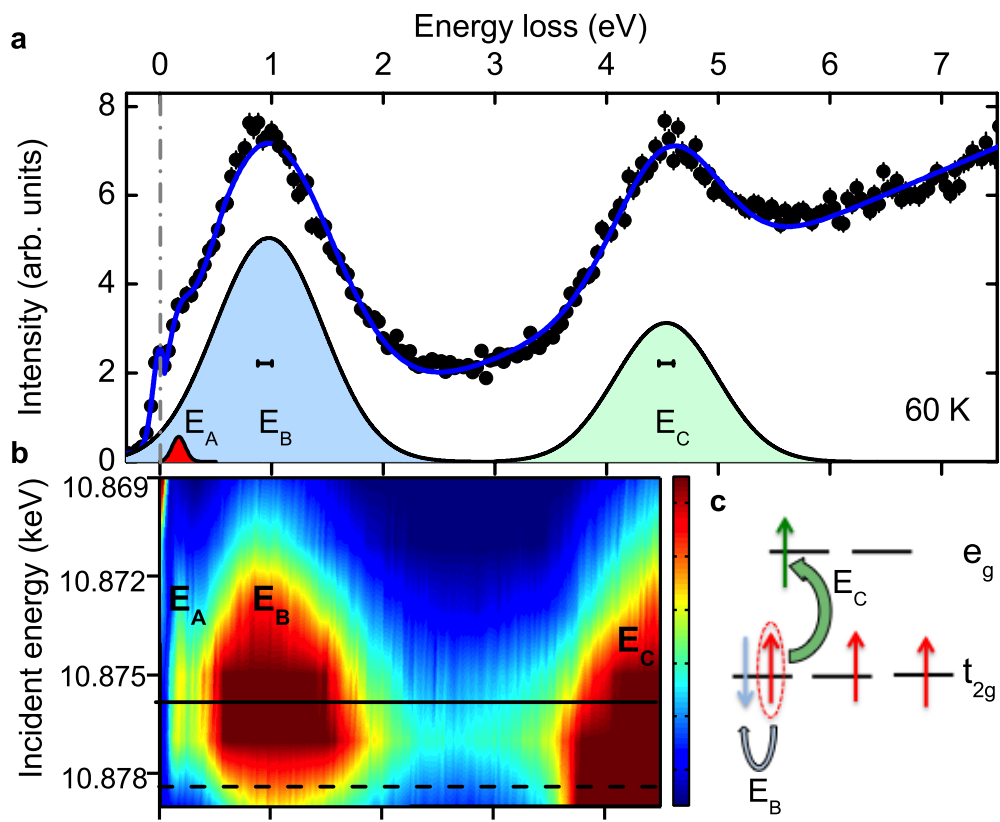

**Figure 2**

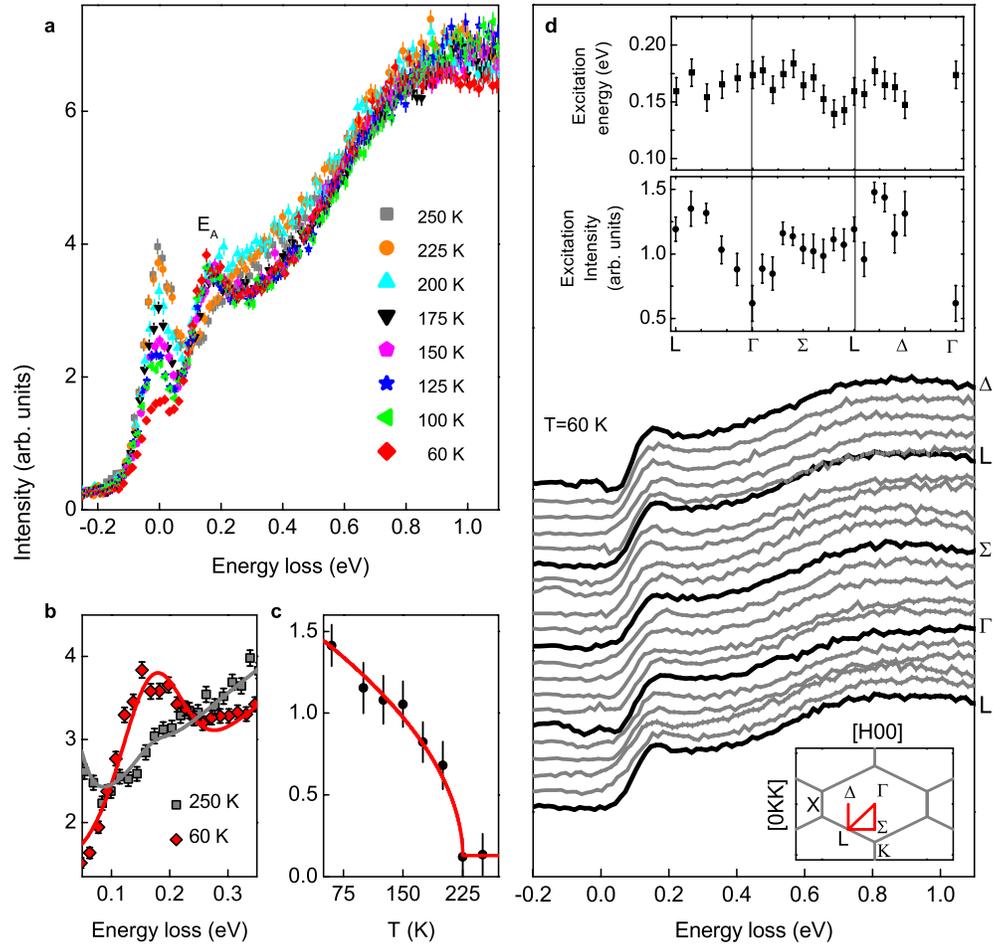

**Figure 3**

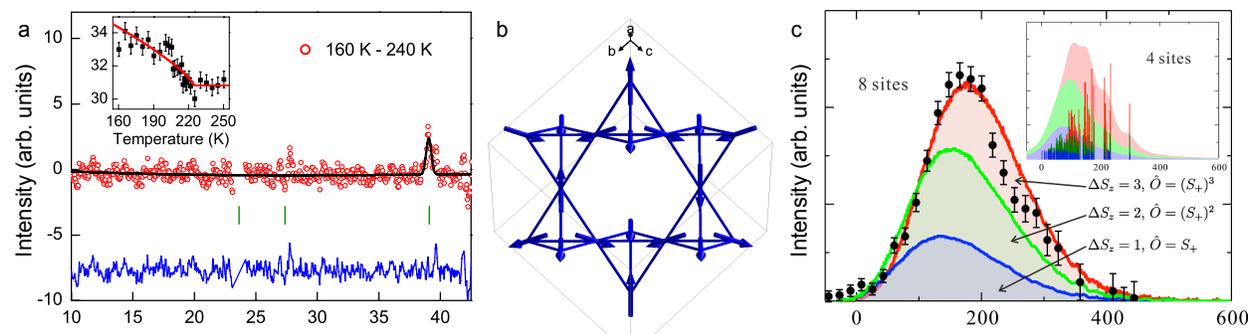

20